\documentclass[reprint, showpacs, superscriptaddress, aps, pre, twocolumn, 10pt]{revtex4-1}
\usepackage{graphicx}
\usepackage{color}
\usepackage{amssymb}
\usepackage{amsmath}
\usepackage{tabularx}
\usepackage{bm}
\usepackage{hyperref}
\usepackage{ltxtable}
\usepackage{natbib} 
\usepackage{longtable}
\usepackage{float}
\newcommand{\la}{\left\langle}
\newcommand{\ra}{\right\rangle}
\newcommand{\be}{\begin{equation}}
\newcommand{\ee}{\end{equation}}
\newcommand{\bse}{\begin{subequations}}
	\newcommand{\ese}{\end{subequations}}
\newcommand{\bea}{\begin{eqnarray}}
\newcommand{\eea}{\end{eqnarray}}
\newcommand{\ba}{\begin{array}}
	\newcommand{\ea}{\end{array}}

\usepackage[usenames,dvipsnames]{xcolor}
\hypersetup{
colorlinks,
citecolor=blue,
filecolor=blue,
linkcolor=blue,
urlcolor=blue}

\begin{document}

\title{Euler Turbulence and thermodynamic equilibrium}
\author{Mahendra K. Verma}
\email{mkv@iitk.ac.in}
\affiliation{Department of Physics, Indian Institute of Technology Kanpur, Kanpur 208016, India}
\author{Shashwat Bhattacharya}
\email{shabhatt@iitk.ac.in}
\affiliation{Department of Mechanical Engineering, Indian Institute of Technology Kanpur, Kanpur 208016, India}
\author{Soumyadeep Chatterjee}
\email{deep@iitk.ac.in}
\affiliation{Department of Physics, Indian Institute of Technology Kanpur, Kanpur 208016, India}

\date{\today}

\begin{abstract}
We perform a unique direct numerical simulation of Euler turbulence using delta-correlated velocity field as an initial condition, and report a  full range of $k^2$ and $k$ energy spectra for 3D and 2D flows respectively, zero energy flux, and Maxwell-Boltzmann distribution for the velocity field.  These are direct verification of the predictions of the absolute equilibrium theory of turbulence.   For a coherent vortex as an initial condition, Euler turbulence transitions from a mixture of nonequilibrium-equilibrium state to a equilibrium state through a process called thermalization.   In this letter, we  present a model for thermalization in Euler turbulence.
\end{abstract}
\maketitle

Physical processes are either in equilibrium  or out of equilibrium~\cite{Zwanzig:book,Livi:book}.  Thermodynamics provides many examples of equilibrium processes, e.g., thermal gas, Bose gas, magnetic systems under heat bath.  In a gas or liquid under equilibrium, apart from fluctuations, there is no net flow of energy or matter from one region to another.  This property is called {\em detailed balance}.    Besides, the average energy and entropy of an equilibrium system remain invariant in time.

On the other hand, nonequilibrium systems are time-dependent with detailed balance broken~\cite{Zwanzig:book,Livi:book}.   Earth's atmosphere, turbulent convection, hydrodynamic turbulence, and earthquakes are some of the examples of such systems.   It turns out that Kolmogorov's theory of turbulence~\cite{Kolmogorov:DANS1941Dissipation, Kolmogorov:DANS1941Structure,Frisch:book,Lesieur:book:Turbulence} for viscous incompressible hydrodynamics provides valuable insights into the nature of nonequilibrium systems.  In this theory,  a viscous fluid is forced at large scales.   The energy injected at the large scale  is transferred to intermediate scale (called inertial range) and then to small scales, where the injected energy is dissipated.  Under a steady state, the inertial-range  energy spectrum is $E(k) = K_\mathrm{Ko} \epsilon^{2/3} k^{-5/3}$, where $ K_\mathrm{Ko}$ is Kolmogorov's constant, $\epsilon$ is the energy flux in the inertial range, and $k$ is the wavenumber.  

In this letter, we focus on turbulence  in incompressible Euler equation, which is the hydrodynamic equation with zero  external force and zero viscosity:    
\be
\partial_t \textbf{u} + (\textbf{u}\cdot \nabla) \textbf{u} = -\nabla p;~~~\nabla \cdot \textbf{u} = 0,
\ee
where \textbf{u} and $p$ are the velocity and pressure fields respectively. 
As we describe below, turbulence in Euler equation, referred to as Euler Turbulence, is very different from  Kolmogorov's model of turbulence, which applies to viscous flows.  Kraichnan~\cite{Kraichnan:JFM1973} and Lee~\cite{Lee:QAM1952} argued that  Euler turbulence has similarities with \textit{equilibrium thermodynamics}, and constructed \textit{absolute equilibrium theory} of Euler turbulence.  By mapping the equations for the Fourier modes of Euler turbulence to those of a Hamiltonian system, and by invoking Liouville's theorem, Kraichnan and Lee derived equilibrium solution of the   three-dimensional (3D) Euler equation with a  finite number of Fourier modes, also called \textit{truncated Euler equations}.  For this solution, the kinetic energy flux vanishes, and the kinetic energy spectrum is $E(k) \sim \gamma k^2/(\gamma^2 - \beta^2 k^2)$, where $\gamma$ and $\beta$ are constants associated with kinetic energy ($u^2/2$) and kinetic helicity (${\bf u \cdot \boldsymbol{\omega}}$) respectively.  Here   ${\bf \boldsymbol{\omega} = \nabla \times u}$ is the vorticity field.   Note that the above equilibrium spectrum is very different from Kolmogorov's $k^{-5/3}$ energy spectrum.  In particular, for nonhelical 3D Euler turbulence ($\beta=0$), $E(k) \sim k^2$, and for two-dimensional (2D) version, the energy spectrum is proportional to $k$~\citep{Kraichnan:ROPP1980}.

There are numerous efforts to verify the aforementioned predictions of Kraichnan~\cite{Kraichnan:JFM1973} and Lee~\cite{Lee:QAM1952}.  \citet{Cichowlas:PRL2005} simulated 3D Euler turbulence using a large-scale Taylor-Green vortex as an initial condition.  For developed turbulence, they observed that $E(k)$ is a combination of Kolmogorov's $k^{-5/3}$ spectrum in the inertial range and $k^2$ at large wavenumbers.  The Taylor-Green vortex induces energy cascade in the inertial range to yield the $k^{-5/3}$ spectrum.  However, the large-wavenumber modes exhibit $k^2$, indicating thermal equilibrium for these modes.   \citet{Cichowlas:PRL2005} claimed that all the Fourier modes would reach equilibrium asymptotically (as time $t\rightarrow \infty$). 

 \citet{Krstulovic:PRE2009} simulated truncated Euler equation with a large-scale helical flow as an initial condition, and obtained Kraichnan's helical absolute equilibrium state at small scales.  Similar behaviour has been observed for truncated Burgers equation~\cite{Majda:PNAS2000,Ray:PRE2011}.   Besides, \citet{Dallas:PRL2015} and  \citet{Alexakis:JFM2019,Alexakis:JFM2020} studied Kolmogorov flow where the forcing is employed at intermediate scales.  They observed that the flow at scales larger than the forcing scale reaches a thermal equilibrium and exhibits $k^2$ energy spectrum.

In  the past numerical simulations, the spectrum predicted by Kraichnan~\cite{Kraichnan:JFM1973} and Lee~\cite{Lee:QAM1952} is not visible for  the whole range of wavenumbers because of the large-scale flow structure employed as an initial condition.   Using a special initial condition, we have been able to achieve the equilibrium configuration for the whole wavenumber range of Euler turbulence.  In our simulation, the velocity field is as random as in thermodynamic gas.  We report these results in the present letter.

In a thermodynamic system under equilibrium, the velocity field is delta-correlated (or uncorrelated), that is,  $\la u_i({\bf r}) u_j({\bf r'}) \ra =( \la u^2 \ra/3) \delta_{ij} \delta({\bf r-r'})$. An   application  of Wiener-Khinchin theorem on the above correlation function yields  equal energy for each Fourier mode, or $E({\bf k}) = \la u^2 \ra/(2 N^3)$, where $N^3$ is the total number of Fourier modes.  The above delta-correlated field  is also called {\em white noise} due to its flat spectrum.  The above modal energy spectrum yields the shell spectrum as $E(k) \sim k^2/\gamma $, where $\gamma$ is a constant.  This relation gets modified in the presence of helicity.  Using helical modes, Kraichnan~\cite{Kraichnan:JFM1973} derived the energy spectrum to be $E(k) \sim \gamma  k^2/(\gamma^2 - \beta^2 k^2)$.

The  connection mentioned above between the delta-correlated velocity field (white noise) and $k^2$ spectrum provides a hint that we should choose white noise as the initial condition for the equilibrium solution of Euler turbulence.     We perform pseudo-spectral simulations~\cite{Boyd:book:Spectral,Canuto:book:SpectralFluid} of 3D and 2D  Euler flows on $64^3$ and $1024^2$ grids respectively, with white noise as an initial condition, and  produce the  equilibrium energy spectra mentioned above.  Note that the external force and viscosity are zero for these flows.  We show below that the above resolutions are sufficient for a demonstration of the equilibrium energy spectrum for Euler turbulence.   Following  \citet{Cichowlas:PRL2005}, we time advance our Euler flow-solver using leap-frog method, which is time-reversible, as well as energy conserving~\cite{Ferziger:book:CFD}.  

As mentioned earlier, we choose white noise as initial condition, contrary to the large-scale flow structures employed by earlier researchers~\citep{Cichowlas:PRL2005,Krstulovic:PRE2009}.  We implement the above random initial condition using Craya-Herring basis~\citep{Craya:thesis,Herring:PF1974} whose unit vectors for a wavenumber ${\bf k}$ are $ \hat{e}_1(\textbf{k}) =( \hat{k} \times \hat{n})/|\hat{k} \times \hat{n}| $  and  $ \hat{e}_2(\textbf{k}) = \hat{k} \times \hat{e}_1(\textbf{k}) $, where $\hat{n}$ is chosen as any direction, and $ \hat{k} $ is unit vector along \textbf{k}.  In this basis, the 3D incompressible velocity field is ${\bf u}(\textbf{k}) = u_1(\textbf{k}) \hat{e}_1 (\textbf{k}) + u_2 (\textbf{k})  \hat{e}_2(\textbf{k}) $, while 2D incompressible velocity field is ${\bf u}(\textbf{k})  =  u_1(\textbf{k})  \hat{e}_1(\textbf{k}) $.

For simulating 3D nonhelical flows (zero kinetic helicity),  we start with $u_1({\bf k})=0$ and $u_2({\bf k}) = \sqrt{2E/N^3} \exp(i \phi_2({\bf k}))$, where $ E = 0.159$ is the total kinetic energy,  $N^3$ is the total number of modes, and the phase $\phi_2({\bf k})$ is chosen to be a random number from uniform distribution in a band of $[0,2\pi]$.     We performed the simulation up to 8 nondimensional time units using leap-frog method and  $dt = 10^{-4}$.  For the 2D simulation, we take $u_1({\bf k}) = \sqrt{2E/N^2} \exp(i \phi_1({\bf k}))$  with $E= 0.0583$, and random phase for $\phi_1({\bf k})$.  We carry out the 2D simulation up to   1 time unit with $ dt = 2.5 \times 10^{-5}$.

During the evolution of the 3D flow, the total kinetic energy $E = 0.15893 \pm 2 \times 10^{-9}$  and the kinetic helicity is  $ 4.03\times 10^{-5} $.  For 2D flow,   $E = (5.832 \pm 0.002) \times 10^{-2}$. Thus, the kinetic energy and helicity are conserved for these flows.    More importantly,  the flow remains random, as in white noise at all times.  Note, however, that the amplitudes and phases of all the modes vary randomly with time.  In Fig. 1(a,b), we exhibit the density plots of the perpendicular vorticity components of horizontal and vertical mid planes. The plots clearly demonstrate the random nature of flow.  

To substantiate the randomness of the flow further, we compute the  probability distribution function ($P(u)$) of the magnitude of the real-space velocity field ($u$) of a snapshot, and test whether it obeys Maxwell-Boltzmann distribution, which is $\sqrt{2/\pi}~a^{-3} u^2 \exp(-u^2/2a^2)$ for 3D, and $a^{-2}u \exp(-u^2/2a^2)$ for 2D, with $a$ as the scale parameter. The numerical $P(u)$'s exhibited in Fig.~2 match quite accurately with the respective theoretical formulas, with $a=0.32$ for 3D and $a=0.24$ for 2D.  Hence, we claim that the velocity field of Euler turbulence is as random as the velocity distribution of  gas molecules in thermodynamic equilibrium.  

\begin{figure}[hbtp]
\includegraphics[scale=0.3]{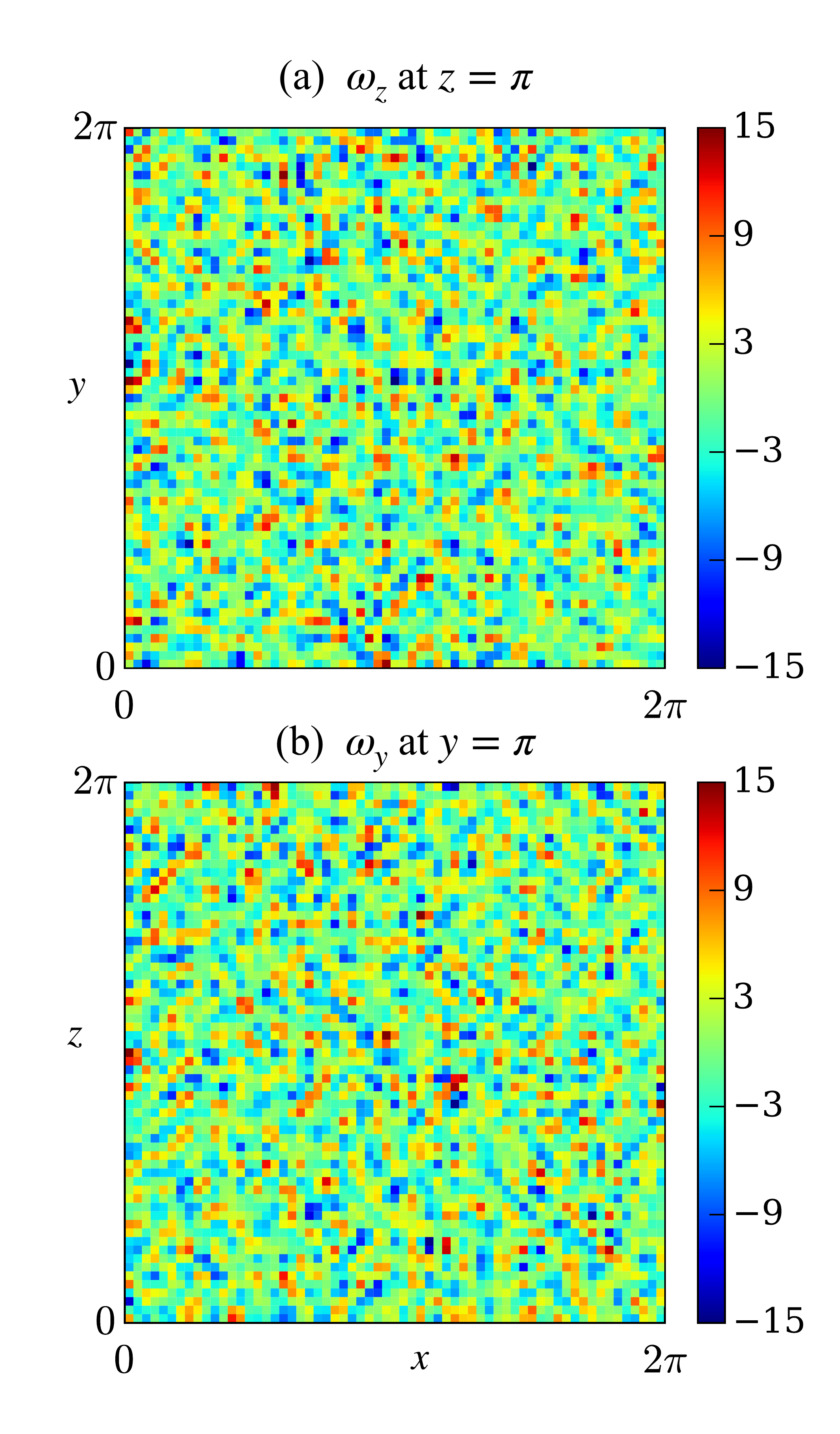}
\caption{For 3D equilibrium flow: Density plots of the perpendicular component of vorticity on (a) horizontal midplane ($z=\pi$), and (b) vertical midplane ($y=\pi$).}
\label{fig:Vorticity}
\end{figure}

\begin{figure}[hbtp]
	\includegraphics[scale=0.5]{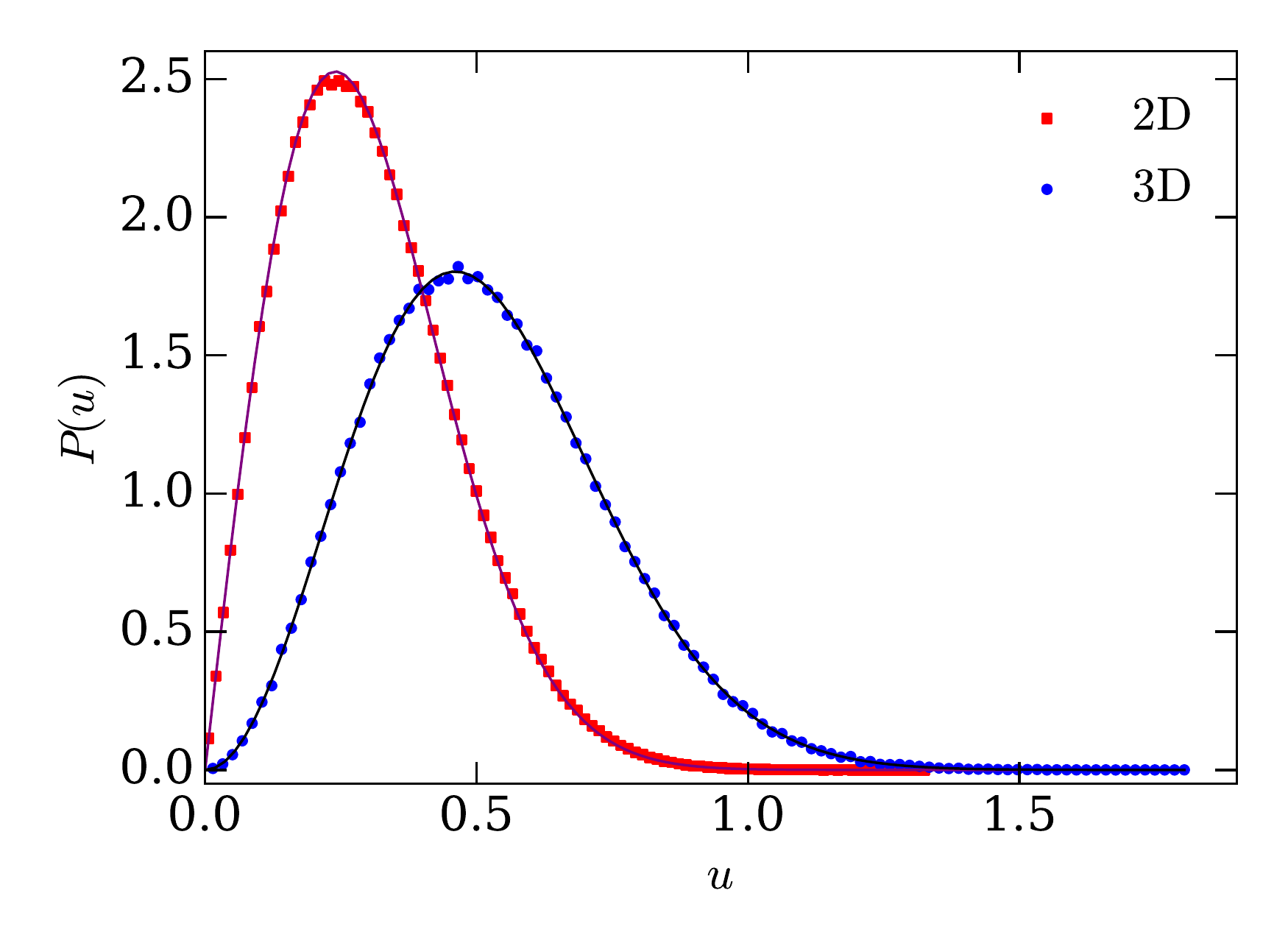}
	\caption{For nonhelical Euler turbulence: Probability distribution functions ($ P(u)$) of the velocity magnitude for 2D (red squares) and 3D (blue circles) real-space flows. The numerical PDFs match closely with Maxwell-Boltzmann distribution for 3D (black curve) and for 2D (purple curve) thermodynamic systems.}
	\label{fig:PDFs}
\end{figure}

\begin{figure}[hbtp]
	\includegraphics[scale=0.24]{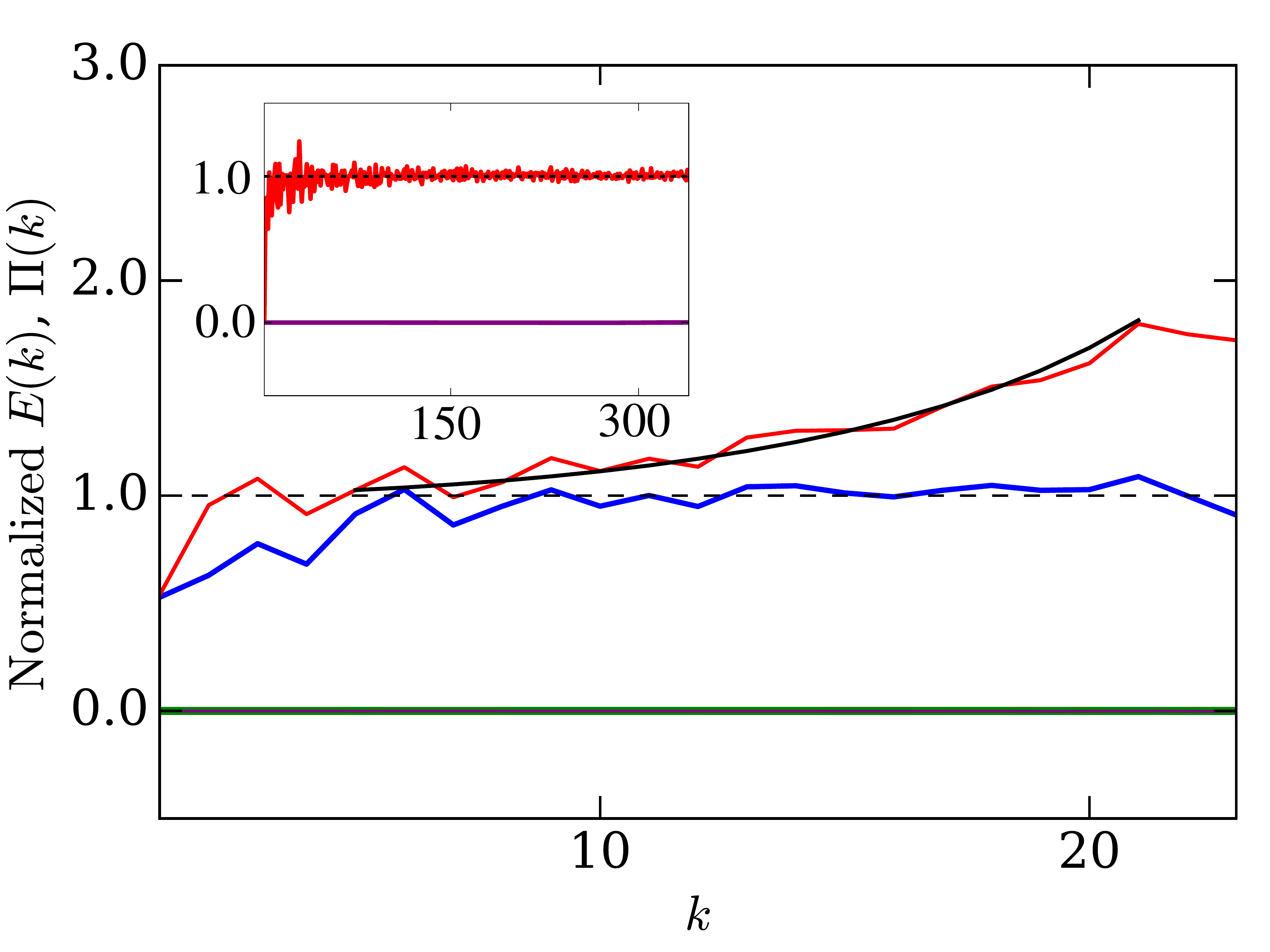}
	\caption{For 3D equilibrium flow: Normalized energy spectrum [$ E(k)/(k^2C')$, blue curve] and flux (green curve) for the nonhelical simulation;   the normalized energy spectrum [$ E(k)/(Ck^2/\gamma^2)$, red curve] and flux (purple curve) for the helical run. The black curve represents $(1-\beta^2k^2/\gamma^2)^{-1}$.  The inset contains the normalized energy spectrum [$ E(k)/(kC_\mathrm{2D})$, red curve] and flux (purple curve) for the 2D simulation.  Here, $C'=2.3\times 10^{-5}, C=2.1\times 10^{-4},   C_\mathrm{2D} = 7.8\times 10^{-7}$. }
	\label{fig:Spectrum_3D}
\end{figure}

We proceed further and compute the energy spectra and fluxes for the two runs.  As shown in Fig.~\ref{fig:Spectrum_3D}, the energy fluxes are zero (apart from fluctuations).  Besides, the normalized energy spectra, $E(k)/(k^2 C')$ for 3D and $ E(k)/(kC_\mathrm{2D})$ for 2D, are flat. Hence we claim that  $ E(k) $ for 3D and 2D  vary as $k^2$ and $k$ respectively for the whole range of wavenumbers.   These results are consistent with the delta-correlated (white noise) nature of the real-space velocity field, thus validating the predictions of absolute equilibrium theory~\cite{Kraichnan:JFM1973, Lee:QAM1952}.    Using field-theoretic arguments, Verma~\cite{Verma:PR2004, Verma:book:ET} has shown that the equipartitioned Fourier modes yield zero kinetic energy flux.  

We perform another numerical simulation to test the effects of kinetic helicity $H$.  For the initial condition of this run, we choose $|u_1({\bf k})| = |u_2({\bf k})| = \sqrt{E/N^3} $, and a random $\phi_2({\bf k})$ from a uniform distribution in $[0,2\pi]$.  The other phase is $\phi_1({\bf k}) = \phi_2({\bf k}) - \sin^{-1}\sigma_c({\bf k})$, where  $\sigma_c({\bf k}) =   \Re[ {\bf u^*({\bf k})} \cdot \boldsymbol{\omega}({\bf k)}]/ (k |\textbf{u}(\textbf{k})|^2) $.  To inject significant kinetic helicity, we choose $\sigma_c(\textbf{k} )= 0.9$ for all \textbf{k}'s.  We employ the same time stepping scheme and $dt$ as in nonhelical case.  The total energy and total kinetic helicity for the run are $0.3179 \pm 0.0001$ and $ 11.817 \pm 0.002  $ respectively; thus, they remain conserved throughout the run.

For the helical run, the real-space velocity field is as random as in Fig.~\ref{fig:Vorticity}.  The energy spectrum, however, deviates at large $k$, as predicted by Kraichnan~\cite{Kraichnan:JFM1973}.  For quantitative match, in Fig.~\ref{fig:Spectrum_3D}, we plot $  E(k)/[Ck^2/\gamma^2]$  vs. $k$.   Our numerical data fits quite well with the  function  $(1- \beta^2 k^2/\gamma^2)^{-1}$, where the best fit parameters are $\gamma=3.14\pm 0.02$ and $\beta=0.100\pm 0.002$.  Since the probability distribution function $P(E,H) \sim \exp(-\gamma E -\beta H)$, from correspondence with statistical physics, we expect that $\gamma = 1/E \approx 3.14$ and $\beta  = 1/H \approx 0.085$.  Thus, the parameters for the best-fit curve are quite close to the above estimates.  Hence, we believe that the velocity field of the helical run too is under equilibrium, as predicted by the absolute equilibrium theory~\cite{Kraichnan:JFM1973}.  As far as we know, this is the first quantitative numerical validation of  the predictions of the absolute equilibrium theory for the whole range of wavenumbers.

In our simulations,  the random initial condition plays a crucial role in yielding the equilibrium configuration.  The earlier 3D runs by \citet{Cichowlas:PRL2005}  using Taylor-Green vortex as an initial condition yields a composite spectrum: $k^{-5/3}$ at small wavenumbers and $k^2$ at large wavenumbers. The simulation by \citet{Krstulovic:PRE2009} using large-scale ABC flow as the initial condition too yields a mixed spectrum. In the flows of  \citet{Cichowlas:PRL2005} and \citet{Krstulovic:PRE2009},  the large-scale vortex induces an energy cascade in the inertial range, thus breaking the detailed balance of energy transfers.  Hence, the Fourier modes corresponding to the large  and inertial scales  are out of equilibrium.  However, the large-wavenumber modes of such flows are in equilibrium.   A comparison of these  numerical simulations with our simulations shows that \textit{ the initial condition plays a critical role in taking a system to equilibrium or  nonequilibrium configuration}. 

\citet{Cichowlas:PRL2005} reported that their numerical results are related to thermalization, which is an important topic of research in nonequilibrium statistical mechanics, both classical and quantum.  They presented a model of thermalization by making certain assumptions on viscous dissipation.  In the following discussion, we formulate a simpler model for thermalization in 3D Euler turbulence.

Let us denote the wavenumber shells in 3D Euler turbulence as $k_0, k_1, ...,k_{N-1}, k_N$, and assume that our initial condition is a large-scale vortex    (with wavenumber $k_0$, as in  \cite{Cichowlas:PRL2005}).  Nonlinear interactions transfer energy from $k_0$ to $k_1$, from $k_1$ to $k_2$, ..., $k_{N-1}$ to $k_N$.  The cascade however stops at $k = k_N$ where the energy piles up.  After sufficiently large  accumulation of energy at $k_N$, the energy starts to grow at  wavenumbers shell   $k_{N-1}$, and then at $k_{N-2}$,  and so on.  This is how the  large wavenumber shells acquire $k^2$ spectrum, as reported by \citet{Cichowlas:PRL2005}.  

\begin{figure}[hbtp]
	\includegraphics[scale=0.9]{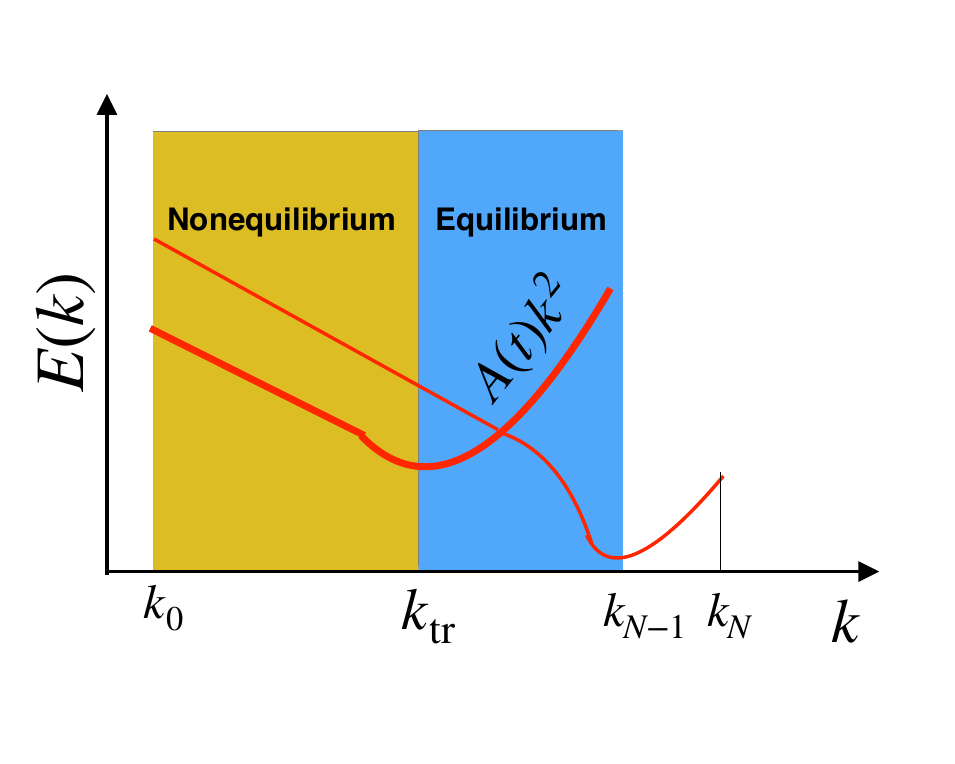}
	\caption{A schematic diagram exhibiting the evolution of energy spectrum $E(k)$ of 3D Euler turbulence during a thermalization process.  The thin red curve represents $E(k)$ during the early phase, while the thick red curve at an intermediate state.  At the transition wavenumber  $k_\mathrm{tr}$, $E(k)$ changes from $k^{-5/3}$ to $k^2$.  The two regimes, nonequilibrium and equilibrium, are represented by yellow and blue colors respectively.   }
	\label{fig:equilibrium}
\end{figure}

Following Kolmogorov's theory of turbulence, the energy cascade rate to the large-wavenumber modes can be estimated as $\epsilon_u = U^3/L \sim U^3 k_0$, where $ L, U$ are the large-scale length and velocity respectively~\cite{Kolmogorov:DANS1941Dissipation, Kolmogorov:DANS1941Structure,Frisch:book,Lesieur:book:Turbulence}.  This energy flux accumulates at large wavenumbers and builds up $ A(t) k^2$ spectrum from the transition wavenumber $k_\mathrm{tr}$, to $k_\mathrm{max}$ (see Fig.~\ref{fig:equilibrium}).  Therefore, in time $t$, 
\be
\epsilon_u t  \sim \int_{k_\mathrm{tr}}^{k_\mathrm{max}} A(t) k^2 dk,
\ee
or
\be
U^3 k_0 t \sim A(t) [k_\mathrm{max}^3 - (k_\mathrm{tr}(t))^3 ].
\label{eq:energetics}
\ee
Over time, $A(t)$ increases, and $k_\mathrm{tr}(t)$ decreases. Using Eq.~(\ref{eq:energetics}) we can deduce the total  time taken for thermalization ($T$) as follows.  During the final stage, $k_\mathrm{tr} \rightarrow k_0 \ll k_\mathrm{max}$ and $A(T) \sim E/N^3$.  Hence,
\be
T \sim \frac{E}{N^3 U^3 k_0} k_\mathrm{max}^3 \sim \frac{L}{U}
\ee
because $k_\mathrm{max} \approx N/2$.  Thus, a 3D Euler flow with large-scale vortex as an initial condition is expected to thermalize in order of one eddy turnover time.  This result is consistent with the estimation of  \citet{Cichowlas:PRL2005}.

Does Euler turbulence have any relevance to realistic flows that have viscosity? The  picture of thermalization presented above provides insights to this question.  A fluid is composed of molecules whose total energy is conserved.  However, we can separate the system into two parts: coherent flow, represented by the yellow region in Fig.~\ref{fig:equilibrium}, and random or thermal flow, represented by the blue region.  During  thermalization, the coherent energy in the inertial range is converted to the thermal energy~\citep{Verma:EPJB2019}.  In the final stage, when all the coherent energy has been converted to thermal energy, the flow reaches an equilibrium.   In the language of statistical mechanics, the yellow and blue regions of Fig.~\ref{fig:equilibrium} could represent \textit{system} and \textit{heat bath} respectively. Thus, the nonequilibrium and equilibrium states of Euler turbulence yield valuable  insights into the process of thermalization.

The above arguments can be extended to quantum systems, at least to superfluids and Bose-Einstein gas.   Many experiments and  numerical simulations of such systems yield Kolmogorov-like $k^{-5/3}$ spectrum~(\cite{Krstulovic:PRL2011,Madeira:ARCMP2020} and references therein) that requires dissipation at small scales.  Small-scale dissipation in such systems are attributed to interactions of condensate with thermal clouds, or to  decay of vortical motion into phonon excitations~(\citet{Barenghi:PNAS2014} and references therein).    This feature may appear odd because quantum systems are energy conserving.  But, the  multiscale energy transfer in Euler turbulence provides an interesting framework to introduce quantum dissipation and thermalization~\citep{Mohsen:book,Weiss:book:QuantumDiss}.    This framework could be an alternative to  other approaches that are typically based on modeling the   interactions between the system and the heat bath (e.g., refer to Caldeira-Leggett model~\citep{Caldera:AP1983}))~\citep{Mohsen:book,Weiss:book:QuantumDiss}.

In summary, we simulate Euler turbulence with delta-correlated velocity as an initial condition and  obtain  equilibrium solution  of Euler turbulence, similar to  thermodynamic equilibrium.  Our final state is very different from those of earlier simulations of Euler turbulence that employ large-scale Taylor-Green vortex as initial conditions.  Thus, the initial condition matters significantly for Euler turbulence simulations.   In addition, in this letter, we present a model for thermalization in Euler turbulence, and argue that the energy transfer framework of  Euler turbulence  and hydrodynamic turbulence could be very useful  for understanding thermalization and dissipation in Hamiltonian systems, both quantum and classical.

The {authors thank}  Stephan Fauve, Marc Brachet, Alex Alexakis, Hal Takasi, Anurag Gupta, Saikat Ghosh, Franck Plunian, and Rodion Stepanov  for useful  discussions.   This work is supported by the project 6104-1  from the Indo-French Centre for the Promotion of Advanced Research (IFCPAR/CEFIPRA).  Soumyadeep Chatterjee is supported by INSPIRE fellowship (IF180094) from Department of Science \& Technology, India.


\begin{thebibliography}{30}
	\expandafter\ifx\csname natexlab\endcsname\relax\def\natexlab#1{#1}\fi
	\expandafter\ifx\csname bibnamefont\endcsname\relax
	\def\bibnamefont#1{#1}\fi
	\expandafter\ifx\csname bibfnamefont\endcsname\relax
	\def\bibfnamefont#1{#1}\fi
	\expandafter\ifx\csname citenamefont\endcsname\relax
	\def\citenamefont#1{#1}\fi
	\expandafter\ifx\csname url\endcsname\relax
	\def\url#1{\texttt{#1}}\fi
	\expandafter\ifx\csname urlprefix\endcsname\relax\def\urlprefix{URL }\fi
	\providecommand{\bibinfo}[2]{#2}
	\providecommand{\eprint}[2][]{\url{#2}}
	
	\bibitem[{\citenamefont{Zwanzig}(2001)}]{Zwanzig:book}
	\bibinfo{author}{\bibfnamefont{R.}~\bibnamefont{Zwanzig}},
	\emph{\bibinfo{title}{Nonequilibrium Statistical Mechanics}}
	(\bibinfo{publisher}{Oxford University Press}, \bibinfo{year}{2001}).
	
	\bibitem[{\citenamefont{Roberto~Livi}(2017)}]{Livi:book}
	\bibinfo{author}{\bibfnamefont{P.~P.} \bibnamefont{Roberto~Livi}},
	\emph{\bibinfo{title}{Nonequilibrium Statistical Physics: A Modern
			Perspective}} (\bibinfo{publisher}{Cambridge University Press},
	\bibinfo{year}{2017}).
	
	\bibitem[{\citenamefont{Kolmogorov}(1941{\natexlab{a}})}]{Kolmogorov:DANS1941Dissipation}
	\bibinfo{author}{\bibfnamefont{A.~N.} \bibnamefont{Kolmogorov}},
	\bibinfo{journal}{Dokl Acad Nauk SSSR} \textbf{\bibinfo{volume}{32}},
	\bibinfo{pages}{16} (\bibinfo{year}{1941}{\natexlab{a}}).
	
	\bibitem[{\citenamefont{Kolmogorov}(1941{\natexlab{b}})}]{Kolmogorov:DANS1941Structure}
	\bibinfo{author}{\bibfnamefont{A.~N.} \bibnamefont{Kolmogorov}},
	\bibinfo{journal}{Dokl Acad Nauk SSSR} \textbf{\bibinfo{volume}{30}},
	\bibinfo{pages}{301} (\bibinfo{year}{1941}{\natexlab{b}}).
	
	\bibitem[{\citenamefont{Frisch}(1995)}]{Frisch:book}
	\bibinfo{author}{\bibfnamefont{U.}~\bibnamefont{Frisch}},
	\emph{\bibinfo{title}{{Turbulence: The Legacy of A. N. Kolmogorov}}}
	(\bibinfo{publisher}{Cambridge University Press},
	\bibinfo{address}{Cambridge}, \bibinfo{year}{1995}).
	
	\bibitem[{\citenamefont{Lesieur}(2008)}]{Lesieur:book:Turbulence}
	\bibinfo{author}{\bibfnamefont{M.}~\bibnamefont{Lesieur}},
	\emph{\bibinfo{title}{{Turbulence in Fluids}}}
	(\bibinfo{publisher}{Springer-Verlag}, \bibinfo{address}{Dordrecht},
	\bibinfo{year}{2008}).
	
	\bibitem[{\citenamefont{Kraichnan}(1973)}]{Kraichnan:JFM1973}
	\bibinfo{author}{\bibfnamefont{R.~H.} \bibnamefont{Kraichnan}},
	\bibinfo{journal}{J. Fluid Mech.} \textbf{\bibinfo{volume}{59}},
	\bibinfo{pages}{745} (\bibinfo{year}{1973}).
	
	\bibitem[{\citenamefont{Lee}(1952)}]{Lee:QAM1952}
	\bibinfo{author}{\bibfnamefont{T.~D.} \bibnamefont{Lee}},
	\bibinfo{journal}{Quart. Appl. Math.} \textbf{\bibinfo{volume}{10}},
	\bibinfo{pages}{69} (\bibinfo{year}{1952}).
	
	\bibitem[{\citenamefont{Kraichnan and Montgomery}(1980)}]{Kraichnan:ROPP1980}
	\bibinfo{author}{\bibfnamefont{R.~H.} \bibnamefont{Kraichnan}}
	\bibnamefont{and} \bibinfo{author}{\bibfnamefont{D.~C.}
		\bibnamefont{Montgomery}}, \bibinfo{journal}{Rep. Prog. Phys.}
	\textbf{\bibinfo{volume}{43}}, \bibinfo{pages}{547} (\bibinfo{year}{1980}).
	
	\bibitem[{\citenamefont{Cichowlas et~al.}(2005)\citenamefont{Cichowlas,
			Bona{\"\i}ti, Debbasch, and Brachet}}]{Cichowlas:PRL2005}
	\bibinfo{author}{\bibfnamefont{C.}~\bibnamefont{Cichowlas}},
	\bibinfo{author}{\bibfnamefont{P.}~\bibnamefont{Bona{\"\i}ti}},
	\bibinfo{author}{\bibfnamefont{F.}~\bibnamefont{Debbasch}}, \bibnamefont{and}
	\bibinfo{author}{\bibfnamefont{M.~E.} \bibnamefont{Brachet}},
	\bibinfo{journal}{Phys. Rev. Lett.} \textbf{\bibinfo{volume}{95}},
	\bibinfo{pages}{264502} (\bibinfo{year}{2005}).
	
	\bibitem[{\citenamefont{Krstulovic et~al.}(2009)\citenamefont{Krstulovic,
			Mininni, Brachet, and Pouquet}}]{Krstulovic:PRE2009}
	\bibinfo{author}{\bibfnamefont{G.}~\bibnamefont{Krstulovic}},
	\bibinfo{author}{\bibfnamefont{P.~D.} \bibnamefont{Mininni}},
	\bibinfo{author}{\bibfnamefont{M.~E.} \bibnamefont{Brachet}},
	\bibnamefont{and} \bibinfo{author}{\bibfnamefont{A.~G.}
		\bibnamefont{Pouquet}}, \bibinfo{journal}{Phys. Rev. E}
	\textbf{\bibinfo{volume}{79}}, \bibinfo{pages}{889} (\bibinfo{year}{2009}).
	
	\bibitem[{\citenamefont{Majda and Timofeyev}(2000)}]{Majda:PNAS2000}
	\bibinfo{author}{\bibfnamefont{A.~J.} \bibnamefont{Majda}} \bibnamefont{and}
	\bibinfo{author}{\bibfnamefont{I.}~\bibnamefont{Timofeyev}},
	\bibinfo{journal}{PNAS} \textbf{\bibinfo{volume}{97}}, \bibinfo{pages}{12413}
	(\bibinfo{year}{2000}).
	
	\bibitem[{\citenamefont{Ray et~al.}(2011)\citenamefont{Ray, Frisch, Nazarenko,
			and Matsumoto}}]{Ray:PRE2011}
	\bibinfo{author}{\bibfnamefont{S.~S.} \bibnamefont{Ray}},
	\bibinfo{author}{\bibfnamefont{U.}~\bibnamefont{Frisch}},
	\bibinfo{author}{\bibfnamefont{S.~V.} \bibnamefont{Nazarenko}},
	\bibnamefont{and}
	\bibinfo{author}{\bibfnamefont{T.}~\bibnamefont{Matsumoto}},
	\bibinfo{journal}{Phys. Rev. E} \textbf{\bibinfo{volume}{84}},
	\bibinfo{pages}{016301} (\bibinfo{year}{2011}).
	
	\bibitem[{\citenamefont{Dallas et~al.}(2015)\citenamefont{Dallas, Fauve, and
			Alexakis}}]{Dallas:PRL2015}
	\bibinfo{author}{\bibfnamefont{V.}~\bibnamefont{Dallas}},
	\bibinfo{author}{\bibfnamefont{S.}~\bibnamefont{Fauve}}, \bibnamefont{and}
	\bibinfo{author}{\bibfnamefont{A.}~\bibnamefont{Alexakis}},
	\bibinfo{journal}{Phys. Rev. Lett.} \textbf{\bibinfo{volume}{115}},
	\bibinfo{pages}{204501} (\bibinfo{year}{2015}).
	
	\bibitem[{\citenamefont{Alexakis and Brachet}(2019)}]{Alexakis:JFM2019}
	\bibinfo{author}{\bibfnamefont{A.}~\bibnamefont{Alexakis}} \bibnamefont{and}
	\bibinfo{author}{\bibfnamefont{M.~E.} \bibnamefont{Brachet}},
	\bibinfo{journal}{J. Fluid Mech.} \textbf{\bibinfo{volume}{872}},
	\bibinfo{pages}{594} (\bibinfo{year}{2019}).
	
	\bibitem[{\citenamefont{Alexakis and Brachet}(2020)}]{Alexakis:JFM2020}
	\bibinfo{author}{\bibfnamefont{A.}~\bibnamefont{Alexakis}} \bibnamefont{and}
	\bibinfo{author}{\bibfnamefont{M.~E.} \bibnamefont{Brachet}},
	\bibinfo{journal}{J. Fluid Mech.} \textbf{\bibinfo{volume}{884}},
	\bibinfo{pages}{87} (\bibinfo{year}{2020}).
	
	\bibitem[{\citenamefont{Boyd}(2003)}]{Boyd:book:Spectral}
	\bibinfo{author}{\bibfnamefont{J.~P.} \bibnamefont{Boyd}},
	\emph{\bibinfo{title}{Chebyshev and Fourier Spectral Methods}}
	(\bibinfo{publisher}{Dover Publications}, \bibinfo{address}{New York},
	\bibinfo{year}{2003}), \bibinfo{edition}{2nd} ed.
	
	\bibitem[{\citenamefont{Canuto et~al.}(1988)\citenamefont{Canuto, Hussaini,
			Quarteroni, and Zang}}]{Canuto:book:SpectralFluid}
	\bibinfo{author}{\bibfnamefont{C.}~\bibnamefont{Canuto}},
	\bibinfo{author}{\bibfnamefont{M.~Y.} \bibnamefont{Hussaini}},
	\bibinfo{author}{\bibfnamefont{A.}~\bibnamefont{Quarteroni}},
	\bibnamefont{and} \bibinfo{author}{\bibfnamefont{T.~A.} \bibnamefont{Zang}},
	\emph{\bibinfo{title}{{Spectral Methods in Fluid Dynamics}}}
	(\bibinfo{publisher}{Springer-Verlag}, \bibinfo{address}{Berlin Heidelberg},
	\bibinfo{year}{1988}).
	
	\bibitem[{\citenamefont{Ferziger and Peric}(2001)}]{Ferziger:book:CFD}
	\bibinfo{author}{\bibfnamefont{J.~H.} \bibnamefont{Ferziger}} \bibnamefont{and}
	\bibinfo{author}{\bibfnamefont{M.}~\bibnamefont{Peric}},
	\emph{\bibinfo{title}{{Computational Methods for Fluid Dynamics}}}
	(\bibinfo{publisher}{Springer-Verlag}, \bibinfo{address}{Berlin Heidelberg},
	\bibinfo{year}{2001}), \bibinfo{edition}{3rd} ed.
	
	\bibitem[{\citenamefont{Craya}(1958)}]{Craya:thesis}
	\bibinfo{author}{\bibfnamefont{A.}~\bibnamefont{Craya}}, Ph.D. thesis,
	\bibinfo{school}{Universit{\'e} de Granoble} (\bibinfo{year}{1958}).
	
	\bibitem[{\citenamefont{Herring}(1974)}]{Herring:PF1974}
	\bibinfo{author}{\bibfnamefont{J.~R.} \bibnamefont{Herring}},
	\bibinfo{journal}{Phys. Fluids} \textbf{\bibinfo{volume}{17}},
	\bibinfo{pages}{859} (\bibinfo{year}{1974}).
	
	\bibitem[{\citenamefont{Verma}(2004)}]{Verma:PR2004}
	\bibinfo{author}{\bibfnamefont{M.~K.} \bibnamefont{Verma}},
	\bibinfo{journal}{Phys. Rep.} \textbf{\bibinfo{volume}{401}},
	\bibinfo{pages}{229} (\bibinfo{year}{2004}).
	
	\bibitem[{\citenamefont{Verma}(2019{\natexlab{a}})}]{Verma:book:ET}
	\bibinfo{author}{\bibfnamefont{M.~K.} \bibnamefont{Verma}},
	\emph{\bibinfo{title}{Energy trasnfers in Fluid Flows: Multiscale and
			Spectral Perspectives}} (\bibinfo{publisher}{Cambridge University Press},
	\bibinfo{address}{Cambridge}, \bibinfo{year}{2019}{\natexlab{a}}).
	
	\bibitem[{\citenamefont{Verma}(2019{\natexlab{b}})}]{Verma:EPJB2019}
	\bibinfo{author}{\bibfnamefont{M.~K.} \bibnamefont{Verma}},
	\bibinfo{journal}{Eur. Phys. J. B} \textbf{\bibinfo{volume}{92}},
	\bibinfo{pages}{190} (\bibinfo{year}{2019}{\natexlab{b}}).
	
	\bibitem[{\citenamefont{Krstulovic and Brachet}(2011)}]{Krstulovic:PRL2011}
	\bibinfo{author}{\bibfnamefont{G.}~\bibnamefont{Krstulovic}} \bibnamefont{and}
	\bibinfo{author}{\bibfnamefont{M.~E.} \bibnamefont{Brachet}},
	\bibinfo{journal}{Phys. Rev. Lett.} \textbf{\bibinfo{volume}{106}},
	\bibinfo{pages}{115303} (\bibinfo{year}{2011}).
	
	\bibitem[{\citenamefont{Madeira et~al.}(2020)\citenamefont{Madeira, Caracanhas,
			dos Santos, and Bagnato}}]{Madeira:ARCMP2020}
	\bibinfo{author}{\bibfnamefont{L.}~\bibnamefont{Madeira}},
	\bibinfo{author}{\bibfnamefont{M.~A.} \bibnamefont{Caracanhas}},
	\bibinfo{author}{\bibfnamefont{F.~E.~A.} \bibnamefont{dos Santos}},
	\bibnamefont{and} \bibinfo{author}{\bibfnamefont{V.~S.}
		\bibnamefont{Bagnato}}, \bibinfo{journal}{Annu. Rev. Condens. Matter Phys.}
	\textbf{\bibinfo{volume}{11}}, \bibinfo{pages}{37} (\bibinfo{year}{2020}).
	
	\bibitem[{\citenamefont{Barenghi et~al.}(2014)\citenamefont{Barenghi, Skrbek,
			and Sreenivasan}}]{Barenghi:PNAS2014}
	\bibinfo{author}{\bibfnamefont{C.~F.} \bibnamefont{Barenghi}},
	\bibinfo{author}{\bibfnamefont{L.}~\bibnamefont{Skrbek}}, \bibnamefont{and}
	\bibinfo{author}{\bibfnamefont{K.~R.} \bibnamefont{Sreenivasan}},
	\bibinfo{journal}{PNAS} \textbf{\bibinfo{volume}{111 Suppl 1}},
	\bibinfo{pages}{4647} (\bibinfo{year}{2014}).
	
	\bibitem[{\citenamefont{Mohsen}(2017)}]{Mohsen:book}
	\bibinfo{author}{\bibfnamefont{R.}~\bibnamefont{Mohsen}},
	\emph{\bibinfo{title}{Classical And Quantum Dissipative Systems}}
	(\bibinfo{publisher}{World Scientific}, \bibinfo{year}{2017}),
	\bibinfo{edition}{2nd} ed., ISBN \bibinfo{isbn}{9789813207936}.
	
	\bibitem[{\citenamefont{Weiss}(1999)}]{Weiss:book:QuantumDiss}
	\bibinfo{author}{\bibfnamefont{U.}~\bibnamefont{Weiss}},
	\emph{\bibinfo{title}{Quantum Dissipative Systems}}
	(\bibinfo{publisher}{World Scientific}, \bibinfo{year}{1999}), ISBN
	\bibinfo{isbn}{9789810240929}.
	
	\bibitem[{\citenamefont{Caldeira and Leggett}(1983)}]{Caldera:AP1983}
	\bibinfo{author}{\bibfnamefont{A.~O.} \bibnamefont{Caldeira}} \bibnamefont{and}
	\bibinfo{author}{\bibfnamefont{A.~J.} \bibnamefont{Leggett}},
	\bibinfo{journal}{Annals of Physics} \textbf{\bibinfo{volume}{149}},
	\bibinfo{pages}{374} (\bibinfo{year}{1983}).
	
\end{thebibliography}

\end{document}